\documentclass[12pt,preprint]{aastex}
\usepackage{graphicx}
\usepackage{epsfig}
\usepackage{multirow}
\usepackage{footnote}

\shorttitle{Study of HE\,1349$-$2305}
\shortauthors{C. Melis et al.}

\begin{document}


\title{Gaseous Material Orbiting the Polluted, Dusty White Dwarf HE\,1349$-$2305}


\author{Carl Melis\altaffilmark{1,8}, P. Dufour\altaffilmark{2}, J. Farihi\altaffilmark{3}, J. Bochanski\altaffilmark{4}, Adam J. Burgasser\altaffilmark{1,4}, S. G. Parsons\altaffilmark{5}, B. T. G\"{a}nsicke\altaffilmark{5}, D. Koester\altaffilmark{6}, Brandon J. Swift\altaffilmark{7}}
\email{cmelis@ucsd.edu}


\altaffiltext{1}{Center for Astrophysics and Space Sciences, University of California, San Diego, CA 92093-0424, USA}
\altaffiltext{2}{D\'{e}partement de Physique, Universit\'{e} de Montr\'{e}al, Montr\'{e}al, QC H3C 3J7, Canada}
\altaffiltext{3}{Department of Physics \& Astronomy, University of Leicester, Leicester LE1 7RH, UK}
\altaffiltext{4}{Massachusetts Institute of Technology, Kavli Institute for Astrophysics and Space Research, Building 37, Room 664B, 77 Massachusetts Avenue, Cambridge, MA 02139, USA}
\altaffiltext{5}{Department of Physics, University of Warwick, Coventry CV4 7AL, UK}
\altaffiltext{6}{Institut f\"{u}r Theoretische Physik und Astrophysik, University of Kiel, 24098 Kiel, Germany}
\altaffiltext{7}{Steward Observatory, University of Arizona, 933 North Cherry Avenue, Tucson, AZ 85721}
\altaffiltext{8}{Joint CASS Departmental Fellow and NSF AAPF Fellow}


\begin{abstract}
We present new spectroscopic observations of the polluted, dusty, helium-dominated
atmosphere white dwarf star HE\,1349$-$2305. Optical spectroscopy
reveals weak Ca\,II infrared triplet emission indicating that metallic gas debris
orbits and is accreted by the white dwarf.
Atmospheric abundances are measured for magnesium and
silicon while upper limits for iron and oxygen are derived
from the available optical spectroscopy. HE\,1349$-$2305 is 
the first gas disk-hosting white dwarf star identified amongst previously known
polluted white dwarfs. Further characterization of
the parent body polluting this star will require ultraviolet
spectroscopy. 
\end{abstract}





\section{Introduction}

White dwarf stars are now known to be polluted by remnant rocky bodies from planetary
systems that otherwise stably orbited their host star while it was on the main
sequence 
\citep[e.g.,][and references therein]{zuckerman07,jura08,farihi09,farihi10b,farihi10a,dufour10,klein10,melis10,zuckerman10,klein11}. 
Prior to being accreted, these rocky bodies are
tidally shredded into disks of dusty material \citep[e.g.,][]{debes02,jura03b}. A subset
of dusty white dwarfs are also host to disks of gaseous metals which
similarly have their origin in the disintegration of remnant rocky bodies from the white dwarf
planetary system
\citep{gaensicke06,gaensicke07,gaensicke08,melis10,farihi11b,brinkworth12}.

\citet{koester05a} and \citet{voss07} describe the atmospheric properties of the
DBAZ (helium-dominated atmosphere with hydrogen and heavy element pollution) 
white dwarf star
HE\,1349$-$2305 (J2000 R.A.\ and Decl.\ of 13 52 44.12 $-$23 20 05.3;
\citealt{epchtein97}). \citet{girven12} detect excess infrared emission toward
HE\,1349$-$2305 indicating that it hosts and accretes from a dusty circumstellar
disk. As a result of parallels between this source and the
potentially water-rich object GD\,61 \citep{farihi11a,jura12}, we obtained
spectroscopic data for HE\,1349$-$2305 to constrain its heavy element
abundances and oxygen content. An unexpected discovery in these
spectroscopic data was the detection of emission lines indicating the
presence of an orbiting gaseous disk. Here we describe
observations of HE\,1349$-$2305 and place it in the context of other
gas disk-hosting white dwarfs.

%% In a manner similar to \objectname authors can provide links to dataset
%% hosted at participating data centers via the \dataset{} command.  The
%% second curly bracket argument is printed in the text while the first
%% parentheses argument serves as the valid data set identifier.  Large
%% lists of data set are best provided in a table (see Table 3 for an example).
%% Valid data set identifiers should be obtained from the data center that
%% is currently hosting the data.
%%
%% Note that AASTeX interprets everything between the curly braces in the 
%% macro as regular text, so any special characters, e.g. "#" or "_," must be 
%% preceded by a backslash. Otherwise, you will get a LaTeX error when you 
%% compile your manuscript.  Special characters do not 
%% need to be escaped in the optional, square-bracket argument.

\section{Observations}
\label{secobs}

\subsection{Nickel Optical Imaging}

Optical imaging was performed on UT 24 March 2010 at Lick
Observatory with the 40" Nickel telescope. These observations
used the facility's Direct Imaging Camera (CCD-C2), a
2048 $\times$ 2048 pixel detector with 15 $\mu$m pixels. The
0.184$^{\prime\prime}$ pixel$^{-1}$ plate scale affords a field
of view of roughly 6.3$^{\prime}$ squared. The detector was
binned by two in rows and columns and was readout in fast mode.

HE\,1349$-$2305 was observed in the $V$-band
\citep{bessell90}. A 4-step dither pattern with 
10$^{\prime\prime}$ steps was repeated with 
60\,second integrations per step position. 
Similar observations were performed for the flux calibrator source Gl\,529 
\citep{bessel90}. Images are reduced by median-combining all
frames to obtain a sky frame and subtracting
this sky frame from each image. Sky-subtracted images are then 
divided by flat field frames obtained by imaging the twilight sky. 
Each science frame is
registered using bright stars in the field and then all science 
frames are median combined to yield the final reduced image.
Detector counts for HE\,1349$-$2305 and Gl\,529
are extracted with an aperture that yields
$\approx$85\% encircled energy (with a negligible correction between
the two sources). This is achieved by extracting counts for both
stars with a 4 binned-pixel (1.5$^{\prime\prime}$) radius circular aperture.
The sky is sampled with an annulus extending from 20-60 pixels.
Uncertainties are derived from the dispersion of measurements
made from individually reduced frames. The uncertainties for
Gl\,529 are propagated into the final quoted uncertainty
for HE\,1349$-$2305.

Nickel
photometry is reported
in Table \ref{tab1349flux} which also includes the DENIS \citep{epchtein97}
$I$-band magnitude and
{\it Galaxy Evolution Explorer} \citep[$GALEX$;][]{martin05} fluxes.

%% In this section, we use  the \subsection command to set off
%% a subsection.  \footnote is used to insert a footnote to the text.


%% This section also includes several of the displayed math environments
%% mentioned in the Author Guide.

\subsection{Gemini Imaging at the Shane 3-m}

Observations of HE\,1349$-$2305 in the $J$-, $H$-, and $K'$-bands 
were performed UT 28 March 2010
with the Gemini Twin-Arrays Infrared Camera \citep{mclean93} mounted
on the 3\,m Shane telescope at Lick Observatory.
We used a 4 position dither pattern with exposure times of 10, 5, and 7\,seconds
with coadds of 15, 30, and 21 per position for each of
$JHK'$, respectively. Total on source integration times of 1800\,seconds were
accrued at each of $JH$ while 3528\,seconds were obtained for $K'$.
The $\approx$3$^{\prime}$ field-of-view of the Gemini instrument enabled simultaneous
observations of two 2MASS \citep{skrutskie06} sources for use in flux calibration.

Data are reduced using in-house IDL software routines. For each filter,
science frames were median
combined to generate a sky-median frame which is then subtracted from each 
science frame. Sky-subtracted frames are then flat-fielded using exposures
of the twilight sky for $JK'$ and the illuminated telescope dome for $H$. Reduced science 
frames are registered with bright point sources within the field. 
Although data were recorded for the longer wavelength chip ($K'$), they are not usable
for accurate photometric measurements due to instrumental difficulties. 
The Gemini photometric results for HE\,1349$-$2305 are reported in
Table \ref{tab1349flux}. $JH$-band results presented herein are consistent with
those presented by \citet{girven12}.

\subsection{MagE Optical Spectroscopy}
\label{sechobs}

Moderate resolution optical spectroscopy of HE\,1349$-$2305 
was obtained on UT 19 March 2011 with the Magellan Echellette 
(MagE; \citealt{marshall08}) mounted on the 6.5\,m
Landon Clay Telescope at Las Campa\~{n}as Observatory. 
%Conditions during the observations 
%were clear with $\sim$1$''$ seeing. 
One 900\,s exposure was obtained 
%at an airmass of 1.14 
with the 0.5$''$ slit aligned with the parallactic angle; this setup provided
3200-10050 \AA\ spectroscopy with a resolving power of $\approx$11000.

Data are reduced using the MASE reduction pipeline \citep{bochanski09}
following standard procedures for order tracing, flat field correction,
wavelength calibration (with ThAr lamp spectra), heliocentric wavelength
correction, optimal source extraction, order stitching, and flux calibration
via observations of Hiltner\,600 \citep{hamuy94}.

\subsection{SpeX Near-Infrared Spectroscopy}

Near-infrared spectroscopy was obtained with
SpeX \citep{rayner03}
mounted on the 3\,m NASA IRTF telescope on UT 22 April 2011.
Prism-mode observations covering 0.8-2.5\,$\mu$m were performed
with a 0.8$^{\prime\prime}$ slit aligned with the parallactic angle. 
Six ABBA nod-patterns were obtained for HE\,1349$-$2305 with 60\,seconds of
integration time and 2 coadds per nod position. A single AB
nod pair of 0.51\,seconds integration time and 10 coadds per nod position
was obtained for the telluric calibration source HD 119752
(A0\,V). Data are reduced with SpeXTool \citep{cushing04,vacca03}.
Absolute flux
calibration of the HE\,1349$-$2305 prism data is accomplished by scaling its 
spectrum to the Gemini $JH$-band measurements.

It is noted that the SpeX data smoothly connect all available near-infrared
photometric data for HE\,1349$-$2305 (Table \ref{tab1349flux} and
\citealt{girven12}).

\subsection{VLT X-shooter Spectroscopy}

Moderate resolution spectroscopy of HE\,1349$-$2305 was obtained
in service mode with X-shooter \citep{dodorico06}
mounted on the 8.2\,m VLT UT2 (Kueyen) telescope. One observation
was obtained on UT 26 May 2011 while two more were obtained
on UT 28 May 2011. UVB arm (3000-5600\,\AA ) observations
were performed with the 0.5$''$ slit resulting in a resolving power of 9900
and exposed for 1475\,seconds per observation. VIS arm (5500-10200\,\AA )
observations were performed
with the 0.4$''$ slit resulting in a resolving power of 18200 and exposed
for 1420\,seconds per observation.
Raw frames are reduced using the X-shooter pipeline version 1.3.7
within \verb+ESOREX+
\begin{footnote}
The ESO Recipe Execution Tool $-$ \verb+http://www.eso.org/sci/software/cpl/esorex.html+;
version 3.9.0 is used.
\end{footnote}.
Standard X-shooter data reduction techniques are employed with
default settings to extract and wavelength calibrate each spectrum.
Relative flux calibration on the science spectrum is performed
with the use of the spectrophotometric
standards LTT\,3218 (May 26th) and EG\,274 (May 28th) 
to derive the instrumental response function. Although
X-shooter coverage extends to the thermal-infrared, data beyond
$\approx$1\,$\mu$m are unusable due to low recorded signal.

\section{Results and Modeling}
\label{secres}

Physical parameters for HE\,1349$-$2305 are adopted from analysis performed
on VLT UVES spectra in \citet{koester05a}
and \citet{voss07}, namely T$_{\rm eff}$ of 18173\,K, log\,$g$ of 8.13 (cgs units), 
and a mass of 0.673\,M$_{\odot}$.
By matching a model white dwarf atmosphere with these parameters
to the observed spectra and photometry we derive a distance
to the white dwarf of 120$\pm$10\,pc (the uncertainty here does not
take into account uncertainties on the white dwarf parameters). Spectral observations 
that extend to wavelengths of $\approx$2.5\,$\mu$m confirm the results of \citet{girven12},
but are not capable of further restricting the dusty disk parameters.
%are consistent with the disk model presented in \citet{girven12}. 
From absorption lines detected in the optical spectra we
derive observed radial velocities (which include contributions from gravitational
redshift and stellar motion) of 40$\pm$30 and 40$\pm$5\,km\,s$^{-1}$ from the MagE and UVES
data, respectively. Difficulties in setting the wavelength zero-point in the X-shooter data
prevent any meaningful radial velocity measurement $-$ these data are corrected to the
white dwarf reference frame by assuming the radial velocity is the same as that measured for the
MagE and UVES data. The contribution from gravitational redshift is estimated to be
35\,km\,s$^{-1}$ and hence the white dwarf systemic motion is $\sim$5\,km\,s$^{-1}$.

\subsection{Abundances}

\begin{figure}
\centering
 \includegraphics[width=130mm,angle=-90]{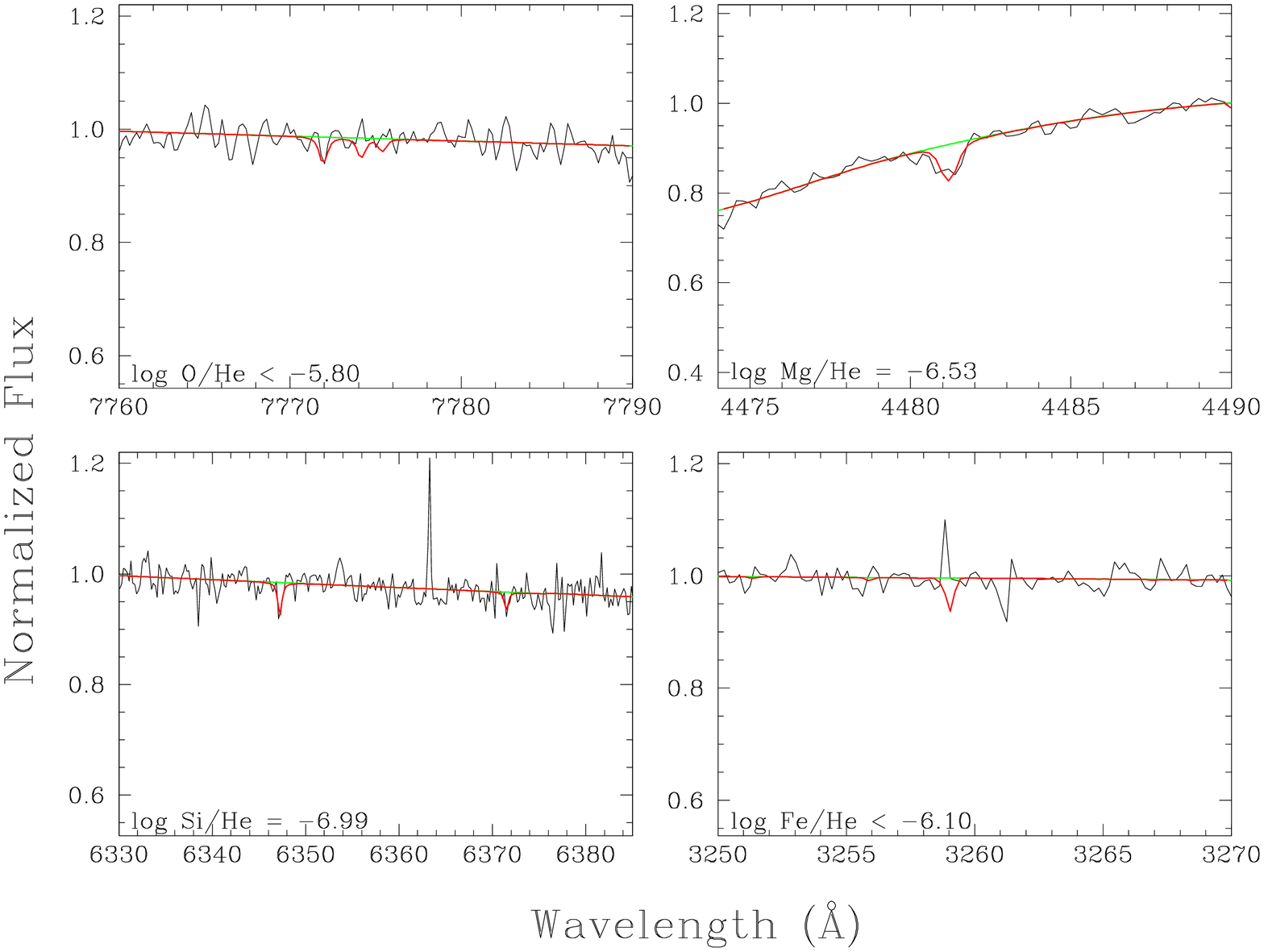}
\caption{\label{figmetals} 
               New abundances and limits from the X-shooter spectra of HE\,1349$-$2305.
               The emission spikes in the spectra are from cosmic rays.
               Magnesium is well-detected, silicon is detected, and iron and oxygen
               are not detected. The limits reported in the panels 
               for oxygen and iron correspond to the absorption
               line strengths shown in their respective panels; the adopted upper limits listed
               in Table \ref{tab1349} are for slightly stronger lines which would have been
               well-detected in the spectrum if they were present. 
               The black curve is the data, the red overplotted curve is the model, and the
               green overplotted curve is the model with the element of interest removed. Wavelengths in
               this figure are presented
               in air.} 
\end{figure}

No absorption lines other than those from H\,I and He\,I are significantly detected in the 
MagE spectrum (Ca\,II H and K lines are marginally detected, but not useful for
abundance modeling). The X-shooter data enable the additional detection of Mg\,II,
Si\,II (Figure \ref{figmetals}), and Ca\,II.
\citet{koester05a} and \citet{voss07} report detections of H\,I and Ca\,II
in their UVES spectra. To try and explore the water content
of the body polluting HE\,1349$-$2305 we calculated upper limits
for the abundances of oxygen and iron, the two major constituents of 
terrestrial rocky minerals not detected (Figure \ref{figmetals}). 
We use a local thermodynamic equillibrium (LTE) 
model atmosphere code similar to that described in 
\citet{dufour05,dufour07}. 
Absorption line data are taken from the Vienna Atomic Line Database
\begin{footnote}
\verb+http://vald.astro.univie.ac.at/~vald/php/vald.php+
\end{footnote}.
We calculate grids of
synthetic spectra for each element of interest. The grids
cover a range of abundances typically from log[$n$(Z)/$n$(He)]= $-$3.0 to
$-$7.0 in steps of 0.5 dex.  We determine abundances or limits
by fitting the expected position of various lines in the spectra
using a similar method to that described in \citet{dufour05}.
Briefly, this is done by
minimizing the value of $\chi$$^2$ which is taken to be the sum
of the difference between the normalized observed and model fluxes over
the frequency range of interest with 
all frequency points being given an equal weight.
Upper limits are derived by comparing model lines of a given abundance
with their expected position in the spectra and determining whether such a line
would be detectable given the local signal-to-noise ratio of the spectrum.

%With the adopted parameters we proceed in fitting any absorption lines 
%detected in the MagE and UVES spectra. 
%We use a local thermodynamic equillibrium (LTE) 
%model atmosphere code similar to that described in 
%\citet{dufour05,dufour07}. 
%Absorption line data are taken from the Vienna Atomic Line Database
%\begin{footnote}
%\verb+http://vald.astro.univie.ac.at/~vald/php/vald.php+
%\end{footnote}.
%We calculate grids of
%synthetic spectra for each element of interest. The grids
%cover a range of abundances typically from log[$n$(Z)/$n$(He)]= $-$3.0 to
%$-$7.0 in steps of 0.5 dex. 
%We then determine the abundance of each
%element by fitting the various observed lines using a similar method
%to that described in \citet{dufour05}. Briefly, this is done by
%minimizing the value of $\chi$$^2$ which is taken to be the sum
%of the difference between the normalized observed and model fluxes over
%the frequency range of interest with 
%all frequency points being given an equal weight. This is
%done individually for each line and the final adopted
%abundances (see Table \ref{tab1349}) are taken to be the average of all the
%measurements made for a given element after removing outliers. 
%Uncertainties are taken to be the dispersion among abundance
%values used in the average.

%We identify contributions from Ca and H in the MagE spectrum
%and in re-analysis of the UVES data presented in \citet{koester05a} and
%\citet{voss07}.
%Upper limits are derived for Si, Mg, O, and Fe. 
All abundance measurements for
HE\,1349$-$2305 are reported in Table \ref{tab1349}.
Abundance measurements for hydrogen and calcium
agree within the errors with those reported in \citet{koester05a} and \citet{voss07}.

\subsection{Gas Emission Lines}

\begin{figure}
 \begin{minipage}[t!]{80mm}
  \includegraphics[width=85mm]{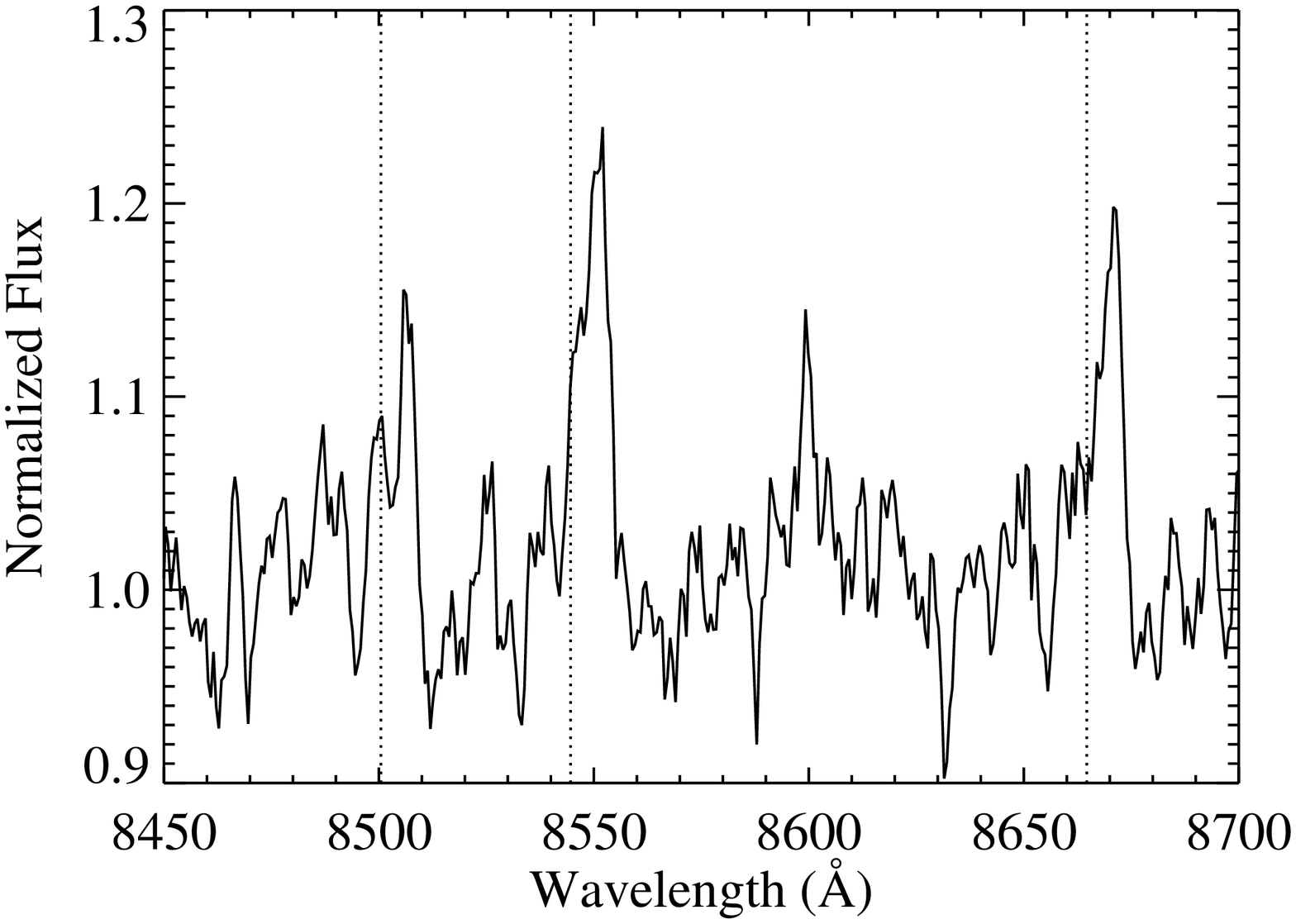}
 \end{minipage}
 \begin{minipage}[t!]{80mm}
  \includegraphics[width=85mm]{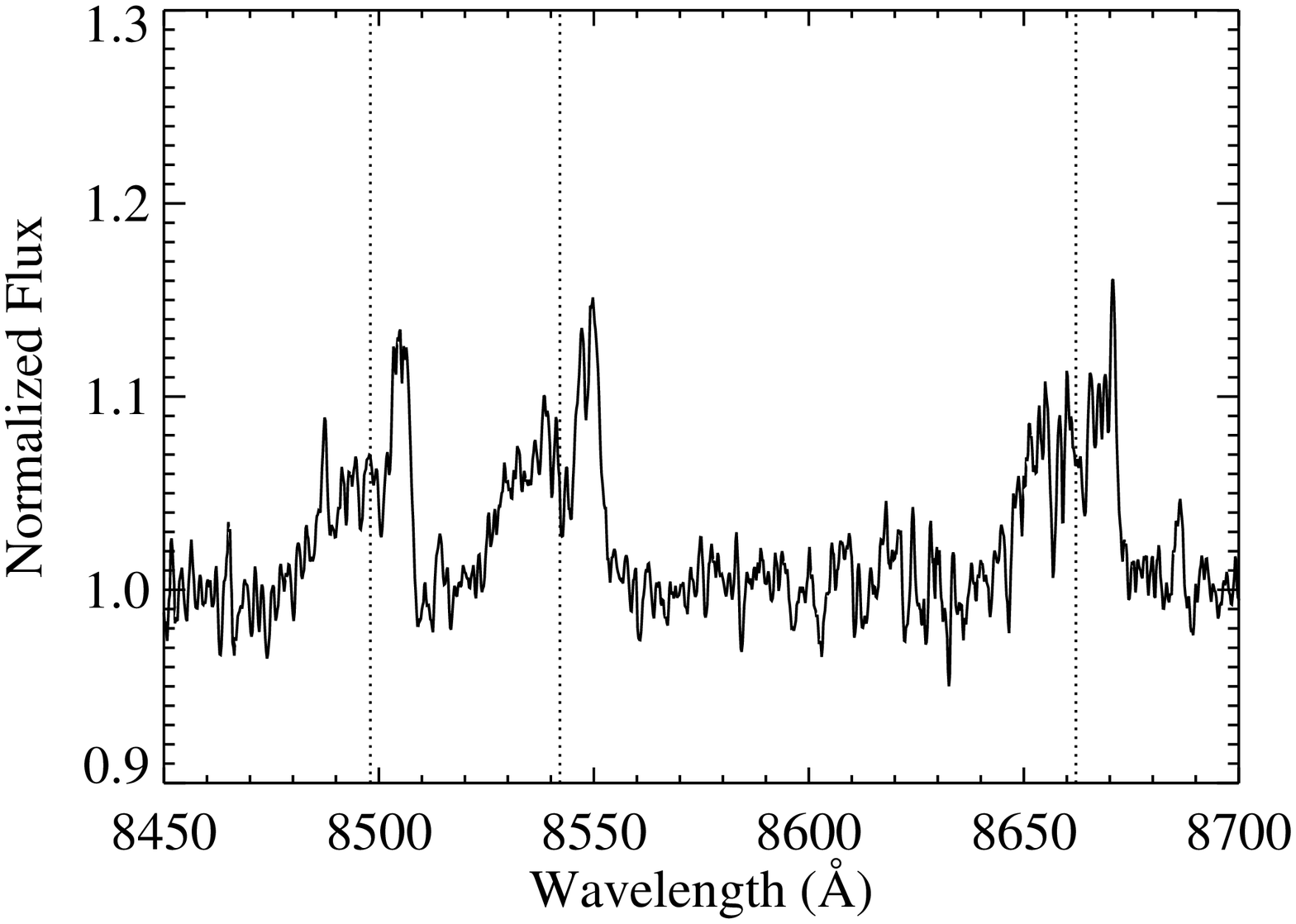}
  \end{minipage}
\caption{\label{figlines} Gas disk emission lines in the MagE (left) and
               X-shooter (right) spectra of HE\,1349$-$2305. The MagE data
               have been smoothed with a 5-pixel boxcar $-$ the low signal-to-noise ratio
               likely prevents detection of the weaker emission hump evident in the X-shooter
               data.
               The vertical dotted lines mark the estimated wavelength 
               position of the white dwarf systemic motion (see Section
               \ref{secres}). Wavelengths in these figures are corrected to
               the heliocentric reference frame and are presented in vacuum
               for the MagE spectrum and air for the X-shooter spectrum.}
\end{figure}

Broad emission lines from the Ca\,II\,infrared triplet (IRT) are detected
in the MagE and X-shooter spectra (Figure \ref{figlines}). For each feature we measure
the maximum gas velocity in the blue and red wings of the detected emission lines,
full velocity width at zero power, and the line flux; these values
are reported in Table \ref{tabgas}. It is not possible to place robust constraints 
on the gas disk inner and outer radii with the available spectra.
Modeling similar to that described in \citet{gaensicke06} suggests that 
the gas disk outer radius is similar to those of the other gas disk-hosting white
dwarf stars, $\sim$100\,R$_{\rm WD}$. From the maximum velocity gas
seen in the disk, and assuming a disk inclination angle of 60$^{\circ}$
(in accordance with inclination angle values used in modeling of white dwarf
dust disks $-$ see e.g., \citealt{farihi09}), we estimate a disk inner radius
of $\sim$15\,R$_{\rm WD}$. If we employ the inclination angle derived
by \citet{girven12} for the dust disk orbiting HE\,1349$-$2305
$-$ $i$$\approx$85$^{\circ}$ $-$ then the inner radius of
the gas disk is slightly higher at $\sim$20\,R$_{\rm WD}$.

More precise constraints on the gas disk structure will require higher signal-to-noise
ratio observations. These observations are likely better carried out using
lower resolution optical spectrographs than those used herein.

%Tinner=1700 K --> Rinner=14 R*
%Touter=  550 K --> Router=63 R*
%
%logg = 8.133 (cgs)
%=> R* = 8.10978*10^8 cm = 0.01166 R_sun
%
%vmax*sini = 780 km/s, vmax = 900.66 km/s
%Rdisk = 1.10126*10^10 cm = 13.57 R*
%
%vmax*sini = 710 km/s, vmax = 819.84 km/s
%Rdisk = 1.32911*10^10 cm = 16.38 R*
%
%vmax*sini = 740 km/s, vmax = 854.47 km/s
%Rdisk = 1.2235*10^10 cm = 15.08 R*
%
%Rdisk(avg) = 15 R*  --> T=1620 K
%
% use Girven et al. 2012 inc=85deg
%vmax*sini = 780 km/s, vmax = 782.97 km/s
%Rdisk = 1.45719*10^10 cm = 17.97 R*
%
%vmax*sini = 710 km/s, vmax = 712.71 km/s
%Rdisk = 1.75869*10^10 cm = 21.68 R*
%
%vmax*sini = 740 km/s, vmax = 742.82 km/s
%Rdisk = 1.61898*10^10 cm = 19.96 R*
%
%Rdisk(avg) = 20 R* --> T=1300 K

\section{Discussion}
\label{secdisc}

HE\,1349$-$2305 is found to host Ca\,II infrared triplet emission lines. The lack
of hydrogen or helium emission lines in the optical and infrared spectra suggest
that this material is metal-rich. Material orbiting the star would be expected
to exhibit a double-peaked emission line morphology similar to that seen for
the other gas disk-hosting white dwarfs \citep{gaensicke06,gaensicke07,gaensicke08,melis10}.
Although only a single peak of emission is significantly detected in the MagE data, a weaker
second peak is evident in the X-shooter data securing the Keplerian origin of
the gas emission. No significant radial velocity shift is evident between the various 
epochs of optical spectroscopy, making the origin of the emission lines from a binary 
companion (or interactions therewith) unlikely. 
As a result, we interpret these emission lines as emanating from a 
gaseous debris disk similar to that described by \citet{gaensicke06}. 
Support of such an interpretation is found through similarities observed for other such 
sources \citep{brinkworth09,melis10} and the detection of thermal-infrared 
excess emission toward HE\,1349$-$2305 indicating that a dusty debris disk also 
orbits the star \citep{girven12}. 
%Although double-peaked line 
%morphologies are usually detected for single
%white dwarfs orbited by metallic gas disks \citep{gaensicke08,melis10}, they are
%not always the rule \citep[e.g.,][]{farihi11b}. It could be possible that HE\,1349$-$2305
%is viewed face-on to our line of sight and hence that no doppler spreading of the
%gas emission is seen. However, from considering the motion of the stellar system
%through space we are able to rule out a face-on disk geometry. In such a case,
%the disk emission would appear as a single line with peak wavelength near that of the 
%white dwarf systemic motion for the given line. 
%As can be seen in Figure \ref{figlines}, this does not appear to be the
%case. We conclude that the gas disk is inclined to our line of sight and
%hence that some double-peaked morphology should be evident.
%It is likely that the quality of the MagE
%spectrum near the Ca\,II infrared triplet lines is not sufficient to detect
%such double-peaked morphology. X-shooter??? Regardless, t
The asymmetry of the 
emission line peak intensities is curious 
(Figure \ref{figlines}). Similar peak intensity contrast 
is evident at a slightly weaker level for SDSS\,J1228 \citep{gaensicke06,melis10} 
and at a slightly stronger level in the 2004 epoch spectrum of Ton\,345 \citep{gaensicke08}.
In regards to the sharp cutoff in line intensity on one edge
of the emission complex and the slower roll-off in intensity at the other edge, 
the emission structure of this
source resembles those of the other gas disk white dwarfs studied at
high spectral resolution \citep{melis10}.
The rough gas disk inner and outer radii are reminiscent
of those measured for the other gas disk-hosting white dwarfs
\citep{gaensicke06,gaensicke07,gaensicke08,melis10}. The outer disk
radius is consistent with the location of the Roche limit for HE\,1349$-$2305,
suggesting that its orbiting gas disk is generated by rocky 
objects that impinge on a pre-existing dusty debris disk (e.g., \citealt{jura08,melis10}).
This is further supported by the fact that the gas disk inner and outer radii
are roughly consistent with the dust disk inner and outer radii
reported in \citet{girven12}.
Thus, despite the inexact characterization of the structure of its orbiting gas disk,
it can be concluded that HE\,1349$-$2305 is being 
polluted by rocky objects from its planetary system similar to other gas 
disk-hosting white dwarf stars \citep[e.g.,][and references therein]{melis10}.

An attempt to constrain the composition of the body polluting 
this star is made by examining the abundances of
the major elemental constituents of rocky minerals: magnesium, silicon, iron, and
oxygen \citep[see e.g.,][and references therein]{klein10}. 
These values are reported in Table \ref{tab1349};
%As discussed in \citet{jura12}, it is difficult to assess the water content of a single
%polluted white dwarf star, especially when abundances are not well constrained.
%This is the case for HE\,1349$-$2305
% for GD61, by number flow O/(Mg+Si+Fe) = 2.17
% for GD61, by number flow O/(Mg+Si) = 2.39
% for HE1349, by number flow O/(Mg+Si) < 7.5 
% M_H = 6.65 x 10^20 g HE1349
% M_H = 2.04 x 10^24 g GD16
% M_H = 5.62 x 10^21 g GD61
from the available data we are unable to make any significant claims about the
water content of the body polluting HE\,1349$-$2305. Comparing the measured
heavy element abundances (calcium, silicon, and magnesium) 
to those of well-studied polluted white dwarf stars suggests that the body
polluting HE\,1349$-$2305 has experienced radiative weathering. 
In particular, the accreted body exhibits the characteristic Si/Mg deficiency and Ca/Mg enhancement
(relative to CI Chondrites) shown by GD\,40 that is interpreted as evidence for
silicate vaporization of the parent rocky body by the intense radiation field from its
evolving host star \citep{klein10,melis11}. These results suggest that HE\,1349$-$2305
could be accreting the remnants of a differentiated rocky body, although tighter
constraints on or detections of oxygen and iron are necessary before solidifying any such claim.
Comparing the pollution of HE\,1349$-$2305 to that of the other gas disk-hosting
white dwarf stars (Table \ref{tab1349}) reveals that HE\,1349$-$2305 is 
among the lowest accretors of the group. Of the helium-dominated atmosphere
white dwarf stars in Table \ref{tab1349}, HE\,1349$-$2305 has a time-averaged
accretion rate that is
lower than the other two (SDSS\,J0738 and Ton\,345) by more than an order of magnitude. 

The identification of a metallic gas disk orbiting HE\,1349$-$2305
shows that there is possibly a significant population of gas disk-hosting white 
dwarf stars that remain to be discovered, perhaps 
even within the currently known population of polluted white dwarf stars.

\section{Conclusions}

We have observed the heavy element and hydrogen polluted white dwarf
HE\,1349$-$2305 in the optical and near-infrared. Our principal results
are that the star hosts Ca\,II infrared triplet emission lines indicative of
orbiting gaseous debris and that there is little evidence of  heavy element
pollution in its optical spectrum (compared to other disk-hosting helium dominated
atmosphere white dwarfs; \citealt{zuckerman07,klein10,klein11,dufour12}).
Observations in the ultraviolet will be necessary
to further examine the elemental composition of the parent body polluting
this white dwarf star.


%% Included in this acknowledgments section are examples of the
%% AASTeX hypertext markup commands. Use \url without the optional [HREF]
%% argument when you want to print the url directly in the text. Otherwise,
%% use either \url or \anchor, with the HREF as the first argument and the
%% text to be printed in the second.

\acknowledgments

%We thank Brandon Swift for performing the Lick Gemini observations. 
%We thank Detlev Koester for kindly providing the reduced UVES data.
C.M.\ acknowledges support from the National Science Foundation under award No.\
AST-1003318. 
P.D. is a CRAQ postdoctoral fellow.
This paper includes data gathered with the 6.5 meter Magellan 
Telescopes located at Las Campa\~{n}as Observatory, Chile. This work
is based partly on observations made with ESO Telescopes at Paranal Observatory
under the program 087.D-0858(A).
This publication makes use of data products from the Two Micron All Sky
Survey, which is a joint project of the University of Massachusetts and the
Infrared Processing and Analysis Center/California Institute of Technology,
funded by the National Aeronautics and Space Administration and the National
Science Foundation. This research has made use of the SIMBAD database
and VizieR service.
Based on observations made with the NASA 
{\it Galaxy Evolution Explorer}. $GALEX$ is operated for NASA by the California 
Institute of Technology under NASA contract NAS5-98034.
This work is supported in part by the NSERC Canada and by the Fund FQRNT (Qu\'ebec).


%% in parentheses, after the keyword, but they will not be verified.

{\it Facilities:} \facility{Magellan:Clay (MagE)}, \facility{VLT:Kueyen (X-shooter)}, \facility{Nickel (Direct Imaging Camera)}, \facility{Shane (Gemini)}, \facility{IRTF (SpeX)}




\clearpage

\clearpage

\clearpage

\begin{deluxetable}{cccc}
\tabletypesize{\normalsize}
\tablecolumns{4}
\tablewidth{0pt}
\tablecaption{Broad-band Fluxes for HE\,1349$-$2305 \label{tab1349flux}}
\tablehead{ \colhead{Band} & \colhead{$\lambda$} & \colhead{$m$\tablenotemark{a}} & \colhead{F$_{\rm obs}$} \\
                      \colhead{}           & \colhead{(nm)}              & \colhead{(mag)}         & \colhead{(mJy)}
}
\startdata
\cutinhead{Lick-Gemini camera\tablenotemark{a} $-$ NIR} \\
$H$  & 1650  & 16.87$\pm$0.08 & 0.19$\pm$0.01 \\
$J$   & 1240  & 16.85$\pm$0.06 & 0.30$\pm$0.02 \\
\cutinhead{DENIS\tablenotemark{a} $-$ Infrared} 
$I$  &  798.2    & 16.70$\pm$0.09 & 0.50$\pm$0.04 \\
\cutinhead{Nickel 40$^{\prime\prime}$\tablenotemark{a} $-$ Optical} 
 $V$ &  544.8    & 16.22$\pm$0.11 & 1.17$\pm$0.12 \\
\cutinhead{$GALEX$\tablenotemark{a} $-$ Ultraviolet} 
$NUV$  &     227.1   &  16.24$\pm$0.03 &   1.16$\pm$0.04 \\
$FUV$  &     152.8   &   16.54$\pm$0.08  &  0.88$\pm$0.06 \\
\enddata
\tablenotetext{a}{Gemini, DENIS \citep{epchtein97}, 
and Nickel magnitudes are on the Vega system.
$GALEX$ measurements are in AB magnitudes.
The $GALEX$ $NUV$ uncertainty is as suggested in \citet{morrissey07}
while the $FUV$ uncertainty is representative of the scatter between the two
separate $GALEX$ detections of HE\,1349$-$2305.}
\end{deluxetable}

\clearpage

\begin{deluxetable}{ccccccccc}
\rotate
\tabletypesize{\normalsize}
\tablecolumns{9}
\tablewidth{0pt}
\tablecaption{Atmospheric Pollution of Gas Disk-hosting White Dwarfs \label{tab1349}}
\tablehead{ 
  \colhead{Star} & 
  \colhead{[H/He]\tablenotemark{a}} &
  \colhead{[O/H(e)]} &
  \colhead{[Mg/H(e)]} &
  \colhead{[Si/H(e)]} &
  \colhead{[Ca/H(e)]} &
  \colhead{[Fe/H(e)]} &
  \colhead{$\dot{M}_{\rm acc,Mg}$\tablenotemark{b}} &
  \colhead{Ref} \\
  \colhead{} &
  \multicolumn{6}{c}{(logarithmic abundances by number)} &
  \colhead{(10$^{8}$\,g\,s$^{-1}$)} &
  \colhead{}
}
\startdata
HE\,1349$-$2305 & $-$4.9$\pm$0.2 & $<$$-$5.6 & $-$6.5$\pm$0.2 & $-$7.0$\pm$0.2 & $-$7.4$\pm$0.2 & $<$$-$5.9 & 1.3 & 1,2,3 \\
% HE 1349 from Voss+07
%Teff=18173 K, logg=8.13, mass of 0.673 M_sun
% M(H) = 3.344 x 10^-13 M_sun = 6.65 x 10^20 g => log(M_He/M_*) = -6.805
SDSS\,J0738 & $-$5.73$\pm$0.17 & $-$3.81$\pm$0.19 & $-$4.68$\pm$0.07 & $-$4.90$\pm$0.16 & $-$6.23$\pm$0.15 & $-$4.98$\pm$0.09 & 146.4 & 4,5 \\
% Teff=13950 K, logg=8.4  M_WD=0.841 M_sun
% log(M_He/M_*) = -6.41, M_CVZ/Mg = 83.33 x 10^21 g, t_set=10^5.258 yrs
SDSS\,J0959 & $-$ & $-$ & $-$5.2 & $-$ & $-$7.0 & $-$ & 0.32 & 6 \\
% Teff=13280 K, logg=8.06, M_WD=0.64 M_Sun
% log(M_H/M_*) = -16.070, M_CVZ/Mg = 1.65 x 10^13 g, t_set= 10^-1.79 yrs
% mdot_Ca = 0.003 x 10^8 g/s
Ton\,345 & $<$$-$4.5 & $-$ & $-$5.2$\pm$0.2 & $-$5.1$\pm$0.2 & $-$6.9$\pm$0.2 & $-$ & 18.3 & 7 \\
% Teff=18525 K, logg=8.28, mass of ~0.7 M_sun
% log(M_He/M_*) = -7.228, M_CVZ/Mg = 3.47 x 10^21 g, t_set=10^4.78 yrs
SDSS\,J1043 & $-$ & $-$ & $-$4.94$\pm$0.24 & $-$ & $-$ & $-$ & 0.73 & 8 \\
% Teff=18330 K, logg=8.09, mass of 0.67 M_sun
% log(M_H/M_*) = -15.794, M_CVZ/Mg = 5.93 x 10^13 g, t_set=10^-1.59 yrs
SDSS\,J1228 & $-$ & $-$ & $-$4.58$\pm$0.06 & $-$ & $-$5.76$\pm$0.08 & $-$ & 2.2 & 9,10 \\
% Teff=22020 K, logg=8.29???, mass of 0.77 M_sun
% log(M_H/M_*) = -15.658, M_CVZ/Mg = 2.13 x 10^14 g, t_set=10^-1.51 yrs
\enddata
\tablerefs{(1) \citet{koester05a}, (2) \citet{voss07}, (3) This work, (4) \citet{dufour10}, (5) \citet{dufour12}, (6) \citet{farihi11b}, (7) \citet{gaensicke08}, (8) \citet{gaensicke07}, (9) \citet{gaensicke06}, (10) \citet{gaensicke08ASPC}.}
\tablenotetext{a}{Hydrogen pollution for helium-dominated atmosphere (DB) white dwarfs. A ``$-$'' in this column indicates that the star has a hydrogen-dominated atmosphere (DA) and that each elemental abundance listed is relative to hydrogen by number. In other columns a ``$-$'' indicates that no measurement exists in the literature.}
\tablenotetext{b}{$\dot{M}$$_{\rm acc,Mg}$=$M_{\rm env,Mg}$/$\tau$$_{\rm diff,Mg}$ where $M_{\rm env,Mg}$ is the mass of magnesium in each star's envelope and $\tau$$_{\rm diff,Mg}$ is the diffusion constant for magnesium (see \citealt{koester09}). For helium-dominated atmosphere
white dwarfs this quantity is averaged over the $\sim$10$^{5}$\,yr settling times.}
\end{deluxetable}

\clearpage

\begin{deluxetable}{lcccc}
\rotate
\tabletypesize{\normalsize}
\tablecolumns{5}
\tablewidth{0pt}
\tablecaption{HE\,1349$-$2305 Emission Line Measurements \label{tabgas}}
\tablehead{
 \colhead{Transition} & 
 \colhead{Equivalent Width\tablenotemark{a}} & 
 \colhead{$v_{max}$sin$i$\tablenotemark{b}} &
 % avg blue edge = -190.03+/-80.83 km/s --- avg right edge = 439.63+/-80.83
 %   RVobs -> blue edge = 230.03+/-86.22 km/s --- RVobs -> red edge = 399.63+/-86.22 km/s
 %       from RVobs of 40+/-30 km/s from MagE HeI lines
 % same for Ton 345
 %  avg blue edge = -648+/-11 km/s --- avg red edge = 621+/-24
 %   RVobs -> blue edge = 684+/-12 km/s --- RVobs -> red edge = 585+/-24 km/s
 %   2.75sigma apart
 \colhead{Full Width\tablenotemark{b}} & 
 % avg full width = 630+/-115 km/s
 \colhead{Total Line Flux\tablenotemark{c}} \\
 \colhead{} & 
 \colhead{(\AA )}           & 
 \colhead{(km s$^{-1}$)} &
 \colhead{(km s$^{-1}$)} &
 \colhead{(10$^{-15}$ ergs cm$^{-2}$ s$^{-1}$)} 
}
%from flux calibrated SpeX data (uses Gemini JH data)
% F_8481 = 0.53 mJy
% F_8544 = 0.53 mJy
% F_8671 = 0.51 mJy
%  take avg => 0.52 mJy for all = 2.16 x 10^-16 ergs/cm2/s/A
\startdata
\multicolumn{5}{l}{\bf MagE $-$ 19 March 2011} \\
%MagE data in VACUUM
Ca\,II $\lambda$8498 & 1.4$\pm$0.3 & $-$190$\pm$140/+410$\pm$110 & 600$\pm$180 & 0.30 \\
% rest = 8498.02 Ang air = 8500.35 Ang vac
%left edge = 8495.0+/-4.0 Ang
%	vle= -188.7+/-140.4 km/s
%right edge = 8512.0+/-3.0 Ang
%	vre= +410.9+/-105.8 km/s
Ca\,II $\lambda$8542 & 2.1$\pm$0.3 & $-$160$\pm$140/+510$\pm$140 & 670$\pm$200 &  0.45 \\
% rest = 8542.09 Ang air = 8544.44 Ang vac
%left edge = 8540.0+/-4.0 Ang
%	vle= -155.8+/-140.4 km/s
%right edge = 8559.0+/-4.0 Ang
%	vre= +510.8+/-140.4 km/s
Ca\,II $\lambda$8662 & 1.7$\pm$0.4 & $-$220$\pm$140/+400$\pm$70 & 620$\pm$160 &  0.37 \\
% rest = 8662.14 Ang air = 8664.52 Ang vac
%left edge = 8658.0+/-4.0 Ang
%	vle= -225.6+/-140.4 km/s
%right edge = 8676.0+/-2.0 Ang
%	vre= +397.2+/-70.6 km/s
\multicolumn{5}{l}{\bf X-shooter $-$ Average of 26 and 28 May 2011} \\
%X-shooter data in AIR (except shifted for figure)
% corr determined from O-line shift of -27.7 km/s and v_helio = -14.93 km/s
%  ...but that leaves a ~15 km/s offset from measured RV, so add +15 km/s
Ca\,II $\lambda$8498 & 1.9$\pm$0.2 & $-$780$\pm$110/+380$\pm$35 & 1160$\pm$120 & 0.41 \\
% rest = 8498.02 Ang air = 8500.35 Ang vac
%left edge = 8475.0+/-3.0 Ang
%      vle= -812.1+/-105.8 km/s
%      vle_corr = -784.3+/-105.8 km/s
%right edge = 8508.0+/-1.0 Ang
%	vre= +352.1+/-35.3 km/s
%      vre_corr = +379.9+/-35.3 km/s
Ca\,II $\lambda$8542 & 1.7$\pm$0.2 & $-$710$\pm$70/+450$\pm$70 & 1160$\pm$100 & 0.37 \\
% rest = 8542.09 Ang air = 8544.44 Ang vac
%left edge = 8521.0+/-2.0 Ang 
%      vle= -740.2+/-70.6 km/s
%      vle_corr = -712.4+/-70.6 km/s
%right edge = 8554.0+/-2.0 Ang
%	vre= +417.9+/-70.6 km/s
%      vre_corr = +445.8+/-70.6 km/s
Ca\,II $\lambda$8662 & 1.6$\pm$0.3 & $-$740$\pm$110/+400$\pm$70 & 1140$\pm$130 & 0.35 \\
% rest = 8662.14 Ang air = 8664.52 Ang vac
%left edge = 8640.0+/-3.0 Ang
%      vle= -766.3+/-105.8 km/s
%      vle_corr = -738.5+/-105.8 km/s
%right edge = 8673.0+/-2.0 Ang
%	vre= +375.9+/-70.6 km/s
%      vre_corr = +403.7+/-70.6 km/s
\enddata
\tablenotetext{a}{Equivalent widths are not corrected for line absorption.}
\tablenotetext{b}{The two different values reported for $v_{max}$sin$i$ correspond to the maximum velocity gas seen in the blue and red wings of the double-peaked emission features, respectively. The blue wing is measured at the continuum blueward of the line while the red wing is measured at the continuum redward of the line. Full velocity width of the emission feature is the velocity extent from the blue to the red wings.}
\tablenotetext{c}{These values are computed by multiplying the reported emission line 
equivalent width measurements by the stellar continuum flux 
(as deduced from the SpeX spectrum) at the emission line location.}
\end{deluxetable}


%% The following command ends your manuscript. LaTeX will ignore any text

\end{document}